\documentclass[12pt]{article}

\usepackage{a4}
\usepackage{amsfonts,amssymb,amsmath,amsthm}
\usepackage{xcolor}
\usepackage{ulem}
\usepackage[colorlinks=true,linkcolor=black,citecolor=teal,urlcolor=blue,filecolor=black]{hyperref}
\usepackage{footnote}
\usepackage{authblk}
\usepackage{float}
\usepackage{physics}

\usepackage{tikz}

\usetikzlibrary{decorations.pathmorphing,backgrounds,calc,shapes, arrows,shapes.misc,shadows}
\tikzset{cross/.style={cross out, draw=black, minimum size=2*(#1-\pgflinewidth), inner sep=0pt, outer sep=0pt},
cross/.default={1pt}}

\newcommand{\eq}[2]{\begin{equation} #1 \label{#2} \end{equation}}

\DeclareMathOperator{\extdm}{d}
\newcommand{\extd}{\extdm \!}

\title{On the dilaton gravity of analogue black holes}

\author[1,2]{P. Castorina}
\author[1,3]{A. Iorio}
\author[1]{J. Kri\v{s}}
\author[4,5,6]{M. Shams Nejati}

\affil[1]{Institute of Particle and Nuclear Physics, Faculty of Mathematics and Physics, Charles University, V Hole\v{s}ovi\v{c}k\'{a}ch 2, 18000 Prague 8, Czech Republic}
\affil[2]{Istituto Nazionale di Fisica Nucleare, Catania branch, Via Santa Sofia 64, 95123 Catania, Italy}
\affil[3]{Department of Physics, Universit\`{a} della Calabria, Via P. Bucci – Cubo 31C, 87036 Arcavacata di Rende (CS), Italy}
\affil[4]{Institute for Theoretical Physics, Technische Universität Wien, Wiedner Hauptstr. 8–10 / 136, 1040 Vienna, Austria}
\affil[5]{School of Physics, Institute for Research in Fundamental Sciences (IPM), P.O.Box 19395-5531, Tehran, Iran}
\affil[6]{Department of Physics, Institute for Advanced Studies in Basic Sciences (IASBS), P.O. Box 45137-66731, Zanjan, Iran}
\begin{document}

\maketitle

\begin{abstract}
  {We investigate whether the gravity models realized by the typical two-dimensional analogue black holes of experimental platforms, such as superconducting quantum circuits, are {indeed} dilaton gravity models{, and if so, their specific classification.}} We identify the most reasonable assumptions these models must satisfy, and determine the dilaton models for which the state-dependence of the Hawking temperature, $T$, can be switched on and off, a feature that is absent in four dimensional black holes. When the analogue black hole exhibits state-independent temperature, as in the cases considered here, the kinematics governing $T$ decouples from the dynamics underlying {the entropy} $S$. Our numerical analysis reveals that the given analogue black holes do not correspond to known dilaton gravity models, limiting their usefulness for extracting theoretical insights. We then show that the logic can be easily {inverted}: starting from established well known dilaton models, one can derive the conditions that laboratory implementations must satisfy. This shifts the challenge from the theoretical perspective to the experimental realization.
\end{abstract}

\section{Introduction}\label{sec:intro}

Feynman’s pioneering ideas on \textit{analogues}~\cite{Feynman} and on \textit{quantum simulations} \cite{Feynman1982}, paved the way to the use of distinct physical systems to describe one another. One road, which starts from there, is that experimentally inaccessible regimes of a given system can be explored, by probing the corresponding accessible regimes of another system, see, e.g., \cite{Xons}. In some cases, this is the only means to experimentally investigate certain high energy exotic scenarios, as their direct probing is impossible.

These are the exciting days when theoretical and experimental advances are rendering those ideas mathematically precise and enabling their practical realization, to the point of progressing toward the full \textit{dynamical equivalence}, see, e.g., the \textit{SYK}\textit{/JT correspondence}\footnote{Sachdev-Ye-Kitaev/Jackiw-Teitelboim} \cite{Sarosi2017}, hence beyond mere \textit{kinematical analogies}.

This paper takes some steps to make the correspondence between the given analogue system and specific \textit{gravity theories} mathematically robust. This is a necessary preliminary step, before further progress can be made, in future works. Those will be the exploration of \textit{quantum architectures} to pursue exact quantum simulations, and then the use of the latter along with the results of this paper, to reproduce the gravity (JT) \cite{JACKIW_ofJT,TEITELBOIM_ofJT} side of the SYK/JT correspondence.

The research in analogue gravity, see, e.g., \cite{Barcelo2005}, was boosted by Unruh’s 1981 paper~\cite{UnruhAnalog}, where he proposed seeking signatures of the Unruh~\cite{bill} and Hawking effects~\cite{haw0} in fluids. Growing theoretical insight and experimental control now enable analogue realizations of key fundamental phenomena and symmetries: Hawking radiation in Bose–Einstein condensates~\cite{MunozdeNova2019}, Weyl symmetry and Hawking/Unruh physics in graphene~\cite{iorio2012,iorio2014,iorio2015,i2,weylgraphene}, anomalies in Weyl semimetals~\cite{Gooth2017}, the duality between laser-driven plasma implosions and supernova explosions~\cite{LOR2001, UniverseLaser}, high-energy heavy-ion collisions~\cite{Castorina2007,caiosa1,caio1} that exhibit hadron production interpretable as a QCD Unruh effect~\cite{Castorina2007} (see also~\cite{grumi} and \cite{casatziorio}).

The focus here is on understanding what sort of gravitational data can  actually be obtained from the analogue system. In particular, we shall deal with two-dimensional (2d) gravity \cite{grumi3,Grumiller:2021cwg}, as many set-ups at the forefront of analogue gravity research, see, e.g., \cite{BHchipNature2023,PachosTanhPRL.130.016701,Linear12018,Linear2Pachos2023,Yang2020}, are in 2d. For instance, in \cite{Yang2020} a general recipe is given on how to realize Hamiltonians, of both fermionic and bosonic matter, on nontrivial (1+1)d gravitational backgrounds, by using 1d quantum circuits.

Those are very important results, nonetheless, the many subtleties of the two dimensions are still to be fully appreciated, and for the analogue gravity model to serve its scope, it is important to spot what sort of gravity model the analogue system actually realizes. Only then the experiments can be used to address questions of theoretical interest. This is significant on its own right, that is why we dedicate this paper to it; but, as mentioned, it is also a necessary preliminary step toward the laboratory realization of JT gravity.

There are actually only few attempts in the literature that exploit the peculiarities of the two dimensions, in this context. Among those, let us point to \cite{grumi} and to \cite{Cadoni:2004my}. For instance, in \cite{grumi}, the exact string black hole (BH), and its limiting case, the Witten BH, were identified as the unique BHs whose near horizon limit coincides with the Rindler spacetime of the QCD Unruh effect of \cite{Castorina2007}. With those results, that hadronic phenomenology can be used to address open issues on the string theory side.

The procedure, as will be explained later in more detail, consists of establishing a certain number of physical assumptions stemming from the analogue physical system, and then, given those assumptions, select the ``best fit''-that is, the model that best fulfills those assumptions.

We shall focus on the typical analogue BH metric, i.e., the one with (squared) lapse function $F(r) \sim \tanh (r)$, obtained in laboratories, see, e.g., \cite{BHchipNature2023}, or studied in theoretical works, see, e.g., \cite{PachosTanhPRL.130.016701}, for different physical setups. We shall, actually, start with the linear approximations of this metric, also studied in the literature, see, e.g., \cite{Linear12018,Linear2Pachos2023}. We shall see that, given the physical setups used to realize such metrics, the Hawking temperature of the dilaton models such analogue systems realize, is actually constant and independent of the ``mass of the hole'', $T =$\,const. This physical fact is behind our assumptions, that guide us toward the identification of the right dilaton model. Notice that our assumptions are different from others made in the literature,  (see, e.g., \cite{Cadoni:2004my}).

This behavior of $T$ is an unusual feature of BH evaporation, due to the 2d. E.g., a Schwarzschild 4d BH notoriously has $T \sim 1/M_{BH}$, while the 3d BH has $T \sim 1/\sqrt{M_{BH}}$, thus their temperature increases as the hole evaporates. Here it remains constant. From this simple example, it is clear that the BH thermodynamics realizable in the lab, with those analogue systems, is peculiar and special, Henceforth, close attention must be paid in order to draw conclusions that hold for the general case.

Furthermore, we shall see that the gravity dilaton model behind the typical analogue system, $F(r) \sim \tanh (r)$, and its linear approximations, is not known nor important for the theory (as it happens, instead, for the QCD case of \cite{grumi}). This is actually a major drawback in an analogue enterprise: what theoretical issues to test with analogue systems that correspond to non-interesting theoretical scenarios?

The paper is organized as follows. In the next Section we discuss some well known issues of 2d gravity, that are crucial for the analysis of this paper, and we put them next to the operational realizations in analogue experiments. In Section \ref{sec:setup} we first explain the procedure to identify the dilaton model, leading to the given metric, and then make our main physical assumptions on the temperature. Section \ref{sec:EOM} is dedicated to write the metric in terms of dilaton quantities, and to make more assumptions, both mathematical and physical. In Section \ref{sec:6} it is discussed why the models of interest are the scale-invariant dilaton models. All the above boils down to write up the ordinary differential equations (ODEs) of Section \ref{sec:masterODE}, that relate the dilaton potential to the analogue metric of interest (we call them the \textit{master ODEs}) and then to establish some general results for the BH thermodynamics of analogue systems in Section \ref{sec:thermo}. With this machinery in our hands, we can eventually face the important concrete cases of Section \ref{sec:results}, where the most demanding computational efforts of this paper have been deployed, when searching for the dilaton models starting from the analogue system (Bottom-up approach). The same Section, however, contains a second part where the opposite approach (Top-bottom) is used. There, the analogue metrics corresponding to important dilaton models are easily obtained and the features distinguishing those models from the ones realized in the lab are addressed. Section \ref{sec:JT} singles out the important case of JT gravity, while the last Section is devoted to our conclusions and outlines further developments.

\section{Preliminaries}\label{sec:preliminaries}

2d gravity exhibits subtleties crucial for theoretical investigations but often obscured by technical details, leading to common misapplications of four-dimensional (4d) BH intuition. A key example is the unwarranted assumption that the state-dependent nature of the temperature of gravitational 4d BHs, $T(M, J, Q)$, directly transfers to 2d analogue BHs. Since our primary assumption concerns the state-dependence of $T$, we address this issue thoroughly here.

{We first clarify the definition of the} ``state'' of a gravitational system, {distinguishing} between 4d and 2d. This distinction is crucial {to our subsequent analysis.}

{In any dimension, {a} ``state'' {refers to} a (classical) field configuration, solving the equations of motion, distinguished by the integration constants of the said equations of motion, which then label physically inequivalent solutions. For BHs, in general, {these} are the mentioned mass $M$, angular momentum $J$ and charge $Q$. In pure 4d gravity, {metric is the only dynamical field; consequently, the metric itself is state-dependent.}}

{{In} 2d dilaton gravity, the {action involves both the metric and the dilaton. Since} the metric carries no dynamical degrees of freedom, it is universal up to diffeomorphisms, {and the state-dependance ia encoded entirely in the dilaton profile. While} one can {employ} coordinate {systems} in which the metric is state-dependent. Nonetheless, if the theory is dilaton gravity, there exists a coordinate transformation that removes such state-dependance from the metric. {Conversely} no coordinate transformation {can} remove the state dependence from the dilaton \cite{grumi3,Grumiller:2021cwg}.}

{{This} contrasts sharply with 4d GR. In 4d the Einstein--Hilbert action is dynamical, and the metric carries propagating degrees of freedom. As a consequence, the integration constants that label a solution (mass $M$, charge $Q$, angular momentum $J$) appear in the metric and cannot be removed by a state-dependent coordinate transformation.}

Let us consider a Schwarzschild BH, whose sates are labeled by a single parameter, $M$. There are two properties of the Schwarzschild BH that hold also in 2d, they do not generally apply to the broader class of 2d BHs relevant for theoretical study \cite{grumi3,Grumiller:2021cwg}, such as those in JT gravity. There is no \textit{priori} reason to assume these properties hold for the analogue BHs under consideration here. These properties are: $(1) \,T$ is inversely proportional to BH mass $M$
\begin{equation}\label{Tschw}
    T = \frac{1}{8\pi} \frac{1}{M} \,,
\end{equation}
in the appropriate units, and that such expression is fixed, once and for all, admitting no coordinate (time) rescaling can alter it.

Indeed, the generic 2d metric in the ``Schwarzschild gauge'' (and for the later convenience also in the Eddington-Finkelstein (EF) gauge) as
\begin{equation}\label{ds2Schw}
    \extd s^2 = - F(r,r_h) \extd t^2 + \frac{1}{F(r,r_h)} \extd r^2
    = 2 \extd v \extd r - F(r,r_h)\,\extd v^2 \,,
\end{equation}
in which $v$ is the ingoing EF time coordinate $v = t + \int dr / F$. For the Schwarzschild BH
\begin{equation}\label{Fschw}
    F_{Schw} (r,r_h) = \left( 1 - \frac{r_h}{r} \right) \,,
\end{equation}
with $r_h = 2 M$. Henceforth,
\eq{
T = \frac{1}{2\pi} \, \kappa = \frac{1}{4\pi} \,\partial_r F(r,r_h)\big|_{r=r_h}  \,,
}{eq:Tsurfgrav}
with $\kappa$ being the surface gravity. The expression \eqref{eq:Tsurfgrav}, with the $F_{Schw}$, and $r_h = 2 M$, leads to \eqref{Tschw}.

A key subtlety in the definition of the surface gravity is its dependence on the normalization of the Killing vector $\chi^{a}$ that generates the horizon.  The defining relation,
\begin{equation} \label{surfgrav}
    \chi^{b}\nabla_{b}\chi^{a} = \kappa \, \chi^{a} \,,
\end{equation}
evaluated on the horizon, {is homogeneous of degree two} in $\chi^{a}$. Under a constant rescaling $\chi^{a}\to\lambda \chi^{a}$ one obtains
\begin{equation}
    \kappa \;\longrightarrow\; \lambda\, \kappa \,,
\end{equation}
indicating that \textit{the surface gravity, and thus the BH temperature, acquires physical meaning only when the normalization of $\chi^{a}$ is fixed by suitable boundary conditions}.

The Killing vector that generates the horizon necessarily includes the time Killing vector\footnote{For a static BH, that is all that is needed, $\chi^a = t^a$, whereas, e.g., for a rotating BH (Kerr) in 4d one has $\chi^a = t^a + \omega_h \phi^a$, etc.} \cite{waldgeneral}. Hence, changing the time coordinate, thereby the normalization of the time Killing vector, may affect the temperature.

This property, however, {does not apply to the BH described by \eqref{ds2Schw} with \eqref{Fschw}}, because the normalization of its Killing vector \textit{is not arbitrary}. The reason is that {this is not a dilaton model and that} the parent 4d spacetime is asymptotically flat, and the time coordinate is fixed by the requirement
\begin{equation} \label{NormMink}
    g_{tt} \;\longrightarrow\; -1 \qquad (r\to\infty) \,.
\end{equation}
This condition selects a \textit{unique} timelike Killing vector corresponding to the proper time of inertial observers at spatial infinity. The reduced 2d theory inherits precisely this normalization: its Killing vector is the dimensional reduction of the 4d Killing vector, and its normalization is fixed by the same asymptotic condition.

With these, the surface gravity, $\kappa$, and the Hawking temperature, $T$, of the Schwarzschild BH are unambiguously defined, both in the 4d Einstein and in the 2d dilaton gravity descriptions\footnote{We shall soon present an argument, independent of asymptotic behavior, that fixes $T$ as a state-dependent quantity for the 2d Schwarzschild BH.}.

In contrast, many intrinsically 2d dilaton gravity models do not admit an asymptotically flat region. Examples include JT gravity (with $\mathrm{AdS}_{2}$ asymptotics), and various string-inspired BHs \cite{grumi3,Grumiller:2021cwg}. In addition, even when asymptotically flat solutions exist, there can be residual freedom in the normalization of the Killing vector. A relevant example is provided by the \textit{dilaton scale invariant models} we discuss later.

We can now turn to the key point for analogue gravity systems. For these setups, the asymptotic region is generally unknown and therefore it should not be used to fix the normalization of the Killing vector in the first place.

In such cases the metric does not approach Minkowski space at large distance, and there is no preferred notion of ``unit time'' at the boundary ($r \to \infty$). As a result, the normalization of the Killing vector $\chi^{a}$ is not fixed by the geometry, and the transformation
\begin{equation}\label{tscale}
    t \;\longrightarrow\; \lambda t
\end{equation}
acts as a genuine global gauge freedom.

Since $\kappa$ scales linearly with $\chi^{a}$, the surface gravity and the corresponding Hawking temperature are defined only up to an overall constant factor unless additional boundary conditions are imposed.

This feature gives an extra freedom to BH temperature of 2d models, as compared to the 4d Schwarzschild BH. In particular, by performing a state-dependent rescaling \eqref{tscale}, one can add or remove the mass dependence from $T$, thereby turning on or off the state-dependence of $T$. We illustrate this mechanism in the case of JT gravity, which is the model of interest for the present discussion.

The standard JT solution, characterized by a mass M and a unit $AdS_2$ radius, corresponds to the metric \eqref{ds2Schw} with
\begin{equation}\label{FJT}
  F_{JT} (r,r_h) = \left( r^2 - M \right) \,,
\end{equation}
which means $r_h = \sqrt{M}$. Since there is no requirement to fix the normalization as in \eqref{NormMink}, we are free to perform
\begin{equation}\label{tildeNorm}
 v  \;\longrightarrow\; \tilde v = \sqrt{M} v \,, \quad r \;\longrightarrow\; \tilde r = r/\sqrt{M} \,,
\end{equation}
yielding the same metric, $\extd s^2 = 2 \extd  v \extd r - \extd v^2 (r^2-M)$, in new coordinates
\begin{equation}\label{dsJTtilde}
  \extd s^2 = 2 \extd \tilde v \extd \tilde r - \extd \tilde v^2 (\tilde r^2-1) \,,
\end{equation}
which means
\begin{equation}\label{FtildeJT}
  F_{JT} (\tilde r, \tilde r_h) = \left( \tilde r^2 - 1 \right) \,,
\end{equation}
with $\tilde r_h = 1$.

It is straightforward to verify that, for the solution \eqref{FJT} the surface gravity is state-dependent, yielding
\begin{equation}\label{TJTSD}
  T = \frac{1}{2 \pi} \, \sqrt{M} \,,
\end{equation}
whereas, for the solution \eqref{FtildeJT} the surface gravity becomes independent of the state, yielding
\begin{equation}\label{TJTSI}
  T = \frac{1}{2 \pi} \,.
\end{equation}

{While the temperature $T$ is a geometric quantity, i.e., it cannot be eliminated through a change of coordinates. The absence of a fixed normalization of the timelike Killing vector in JT gravity allows us to ``scale away'' the state-dependence of $T$.} In the $(v,r)$ coordinates, $T$ is state-dependent; in the $(\tilde v, \tilde r)$ coordinates, $T$ is state-independent. This distinguishes the 2d case from the ``standard astrophysical'' 4d Schwarzschild BHs where the normalization of the Killing vector is fixed by boundary conditions at infinity. Recognizing this key difference is essential when dealing with 2d BH analogues.

Furthermore, the argument that analogue gravity systems generally lack a precisely known asymptotic region -making the normalization of the Killing vector ambiguous- is also applicable to the 2d Schwarzschild solution \eqref{ds2Schw} with the metric \eqref{Fschw}. However, an additional argument applies here, which is crucial for what follows.

In fact, no rescalings
\begin{equation}\label{tildeNormGeneral}
 v  \;\longrightarrow\; \tilde v = M^A v \,, \quad r \;\longrightarrow\; \tilde r = M^B r \,,
\end{equation}
can render the entire 2d Schwarzschild metric state-independent. Some dependence on $M$ remains in either the $dv dr$ term or the 1 or the $-2M/r$ term will have some $M$-dependence. Therefore, while one can find scalings that make $T$ state-independent (e.g., $B = 1 + 2 A$), they would never remove the state-dependence from the Schwarzschild metric. Consequently, \textit{such metrics cannot form the state-space of 2d dilaton gravity models}.

In other words, while one can always make some state-dependent diffeos \eqref{tildeNormGeneral} that eliminate the state dependence of $T$, the resulting structure generally lacks an interpretation as the state-space of a consistent 2d dilaton gravity model. The only exception known so far are dilaton scale invariant models, which will be discussed later.

Conversely, the entropy $S$ -which is arguably the most important thermodynamic quantity besides temperature- \textit{is invariably state-dependent}, regardless of the spacetime dimensionality, the specific gravity theory, or the chosen coordinate system. For instance, in 2d dilaton gravity, the entropy is always given by the coupling constant of the model $k$, times the dilaton evaluated at the horizon, $X_h$,
\begin{equation}\label{Sgeneral}
  S = k \, X_h \,.
\end{equation}
This expression clearly indicates that the thermodynamics relations, such as $T(S)$, depend crucially on the state-dependence of $T$.

What remains to be discussed in this Section, is whether we are truly free to scale the coordinate Killing vector, $\partial_t$, i.e., if we are free to scale the laboratory time. In fact, in analogue systems, coordinates have a special significance: the laboratory realization of a metric in a given coordinate frame is not equivalent to the realization of the same metric in a different coordinate frame. See for example \cite{iorio2014} for relevant discussions. A certain freedom persists, because experimental results obtained in the lab for a given metric -say in the coordinates $(v, r)$-, can then be mathematically transformed via symmetries such as diffeomorphisms (diffeos), conformal or Weyl transformations. These transformations produce a new metric, (for instance in the coordinates $(\tilde v, \tilde r)$), and the results remain valid within the transformed coordinate frame, even if the actual laboratory setup does not physically realize (or cannot realize) the transformed metric. A 3d example, for an ultra-static metric and Weyl symmetry, can be found in \cite{iorio2014}.

These general considerations regarding the physical realization of coordinates in the lab, provide specific insights into the coordinate-dependent state-dependence nature of the analogue BH temperature $T$ as follows: \textit{It is the state-dependence of the physical quantity that plays the role of the analogue BH temperature, $T$, that dictates what coordinates have actually been realized.}

To illustrate with the JT gravity example: if the physical quantity corresponding to the analogue BH temperature $T$, is state-independent, then the JT system is described in the $(\tilde v, \tilde r)$.  In this case, given the expressions in \eqref{tildeNorm}, the time $\tilde v$ must be identified with a \textit{rescaled lab time}. The rescaling factor is state-dependent proportional to $\sqrt M$ and its precise value is determined by the specific physical set-up employed to realize the JT gravity system. Conversely, if the physical quantity that corresponds to the analogue BH temperature $T$, is \textit{state-dependent}, then the JT system is naturally described within the $(v,r)$ frame. A direct consequence is that the time $v$ represents the lab time without any further requirements.

\section{The setup: logic, main assumption and conventions}\label{sec:setup}

We can now focus on the typical 2d analogue BHs of \cite{BHchipNature2023,PachosTanhPRL.130.016701,Linear12018,Linear2Pachos2023}. We shall look for the dilaton gravity model they correspond to. The logic we shall follow is:

{\it
The analogue system provides a 2d metric given by
\eq{
\extd s^2=2\extd v\extd r - F(r,r_h)\,\extd v^2 \,,
}{eq:ana0}
in EF gauge, in which $F$ is an arbitrary function and $r_h$ is an integration constant. Then, we make the necessary physical and mathematical assumptions, and obtain the dilaton potential function \cite{grumi3,Grumiller:2021cwg}, ${\cal V} (X)$, such that all solutions of the dilaton equations of motion are given by \eqref{eq:ana0}. This identifies the dilaton model behind that analogue metric, that is our goal.}

We are interested in stationary 2d metrics as the simplest analogue models of BHs, and the metric \eqref{eq:ana0} represents the most general metric within this class, up to diffeos. This freedom to perform diffeos must be exercised with consideration for the special status of coordinates in the laboratory, as discussed in the preceding Section.

Furthermore, we seek stationary BH solutions at finite Hawking temperature and hence there must be at least one non-extremal Killing horizon. Thus, the function $F$ has to have at least one zero and it must be a single zero. Without loss of generality, we assume the single zero of $F$ is located at $r=r_h$, that is the horizon.

The central physical assumption we make, based on the laboratory realizations of such BHs, is that \textit{the temperature, which can generally depend on the physical state, is actually state-independent}.

As discussed in the previous Section, although this might come as a surprise, for stationary BHs in 4d general relativity, where $T(r_h)$ depends on the state characterized by $r_h(M,Q,J)$, this is not the case in 2d. In this context, we must evaluate the state-dependence of both $T$ and $r_h$ on a case-by-case basis, allowing for all four possible combinations of their state-dependence.

For specific examples, consider \cite{BHchipNature2023} and \cite{PachosTanhPRL.130.016701}, in which $F (r) = \xi f(r)$, with $f(r) = \tanh (r)$. The state-dependence of temperature can be expressed as
\eq{
T = \frac{\xi}{4\pi} \,\partial_r f(r)\big|_{r=r_h} = \frac{\xi}{4\pi} \,,
}{eq:Tana}
Here, state-dependence of temperature is determined by the state-dependence of $\xi$. Similar considerations hold for the linear approximation discussed in \cite{Linear12018,Linear2Pachos2023}.

From a mathematical perspective, we have the freedom to perform a coordinates rescaling in \eqref{eq:ana0}:  $v \to v / \sqrt{\xi}$ and $r \to r \sqrt{\xi}$, which allows us to set $\xi=1$. However, it is essential to recognize that $\xi$ represents the experimental quantities used in the lab to reconstruct the BH metric. For instance, the authors of \cite{BHchipNature2023} denote this parameter as $\beta$ (related to the transmons' constant frequency), while the authors of \cite{PachosTanhPRL.130.016701} refer to it as $\alpha$ (related to the coupling constant of the spin-chain). Therefore, it is crucial not to set $\xi$ to one, as this would obscure the crucial discussion on state-dependence of $T$.

Before proceeding, \textit{we make the important assumption that $r_h$ is state-dependent}. This is a natural assumption in any gravity theory, as it is physically intuitive that changes in BH's size correspond to distinct physical states. In analogue systems, for instance, the setup described in \cite{BHchipNature2023} implies changing the location, $j_h$ of the transmon coupler $C_j$, where the couplings between transmon qubits change sign. Whereas, in the setup of \cite{PachosTanhPRL.130.016701}, this corresponds to changing the location, $n_h$, of the spin within the spin-chain, where the interaction between spins changes sign. Consequently, we can consider $r_h$ as a label for these physical states (see later here, too). Here $j= 1,2,...$ is the discrete label that counts the qubits and their couplers, while $n= 1,2,...$ is the discrete label that counts the spins in the chain.

We now present physical arguments for the state-independence of the temperature of the typical analogue BHs. In \cite{BHchipNature2023} the temperature $T$ does not depend on the location of the horizon but only on a constant $\xi$ (where $\xi / (2 \pi) \simeq 4.39$MHz in that work). Henceforth, the temperature of this BH cannot inherit the state-dependence of $j_h$ (which corresponds to $r_h$ in the continuum version). Similar reasoning holds for the $T$ of the spin-chains described in \cite{PachosTanhPRL.130.016701}, where the temperature is determined by a constant $\xi$ (denoted $\alpha$ in that reference), independent of the horizon's location.

This implies that as the BH grows or shrinks, even though the number of its physical states change, $T$ does not change. This observation strongly supports the state-independence of $T$.

It is worth noting that the analogue BHs discussed in \cite{BHchipNature2023,PachosTanhPRL.130.016701} do not dynamically change their `size' through processes like evaporation or back-reaction. Their size, specifically the location of $j_h$ or $n_h$, is a quantity that can only be statically engineered for each given experimental setup. Nevertheless analogue BHs with different `sizes' (i.e., different $j_h$ or $n_h$) all exhibit the same temperature proportional to $\xi$.

On the other hand, as previously recalled, \textit{entropy by its very same definition, is inherently a state-dependent quantity}. As a result, the temperature which is determined entirely by the metric in any gravity theory, and the entropy, which is determined entirely by the dilaton, as is standard in dilaton gravity theories \cite{grumi3,Grumiller:2021cwg}, are independent quantities.

Other approaches, such as those presented in ~\cite{Cadoni:2004my}, presuppose knowledge of the entropy within the line-element, thereby leading to different classes of 2d dilaton gravity models. We consider it implausible to postulate the correct entropy, - a dynamical quantity -, within the context of an analogue BH background metric, which is a kinematical quantity. While it is possible that no 2d dilaton gravity description exists for these systems, if such a description is assumed to exist, our proposed approach represents the most reasonable path forward.

Motivated by the cases of interest, $f(r,r_h)=\tanh(r-r_h)$ as in \cite{BHchipNature2023,PachosTanhPRL.130.016701} and $f(r,r_h) \propto (r-r_h)$ as in \cite{Linear12018,Linear2Pachos2023}, we make our most restrictive mathematical assumption, i.e., that\footnote{Since $r_h$ denotes the position of a Killing horizon, the function $f$ cannot be arbitrary but must vanish at $r=r_h$, and this zero must be simple for non-extremal horizons. If  $r=r_h$ represents the only horizon, then  $f$ must have no other zeros. This condition significantly constrains the constrains the admissible form of $f$. Note also that $f$ can have a single zero at the horizon, $r_h$, for $f(r,r_h) = (r - r_h) g(r,r_h)$, with $g(r,r_h) \neq g(r - r_h)$, e.g., $g(r,r_h) = e^{r/r_h}$.} $f(r,r_h) = f(r - r_h)$. Under this assumption, the general analogue metric we consider takes the form
\eq{
\extd s^2 = 2 \extd v\extd r - \xi \, f(r - r_h)\,\extd v^2 \,.
}{eq:ana3}

Before introducing the dilaton language employed in the remainder of this paper, we briefly revisit the state-dependence of temperature $T$ for these models focusing on the metric
\eq{
\extd s^2 = 2 \extd v\extd r - \xi \, \tanh(r-r_h) \,\extd v^2 \,.
}{eq:ana3tanh}
The analysis closely parallels that of the JT BH discussed in Section \ref{sec:preliminaries}. Performing
\begin{equation}\label{tildeNormtanh}
 v  \;\longrightarrow\; \tilde v = r_h v \,, \quad r \;\longrightarrow\; \tilde r = r/r_h \,,
\end{equation}
we obtain the metric
\eq{
\extd s^2 = 2 \extd \tilde v \extd \tilde r - \frac{\xi}{r^2_h} \, \tanh\left( r_h (\tilde r - 1) \right) \,\extd \tilde v^2 \,,
}{eq:ana3tanhtilde}
which leads to
\eq{
T = \frac{\xi}{4\pi} \, \frac{1}{r_h} \,,
}{eq:TanaSD}
Since $r_h$ is assumed to be state‑dependent, it follows that the temperature is state‑dependent as well.

The key question is therefore which coordinate frame corresponds to the laboratory realization: $(v,r)$ or $(\tilde v, \tilde r)$. As discussed in Section \ref{sec:preliminaries}, the operational definition of temperature determines this choice. The previous discussion on the lab realization of such BHs clearly indicate that the measured temperature $T$ is state-independent, implying that the correct laboratory frame is $(v,r)$. It is also worth noting that the two BH metrics with linear $f$ introduced in \cite{Linear12018,Linear2Pachos2023} are related by precisely the transformation \eqref{tildeNormtanh}.

A complementary viewpoint can be obtained by considering the more general metric \eqref{eq:ana3}, and simply shifting the radial coordinate, by redefining $\hat r = r-r_h$. In this form, both the metric and the function $f$ become independent of $\hat r_h = 0$
\eq{
\extd s^2 = 2 \extd v\extd\hat r - \xi \, f(\hat r)\,\extd v^2 \,.
}{eq:ana3SI}
Under these assumptions, if $\xi$ (whose properties are fixed by the phenomenology) is state-dependent, the surface gravity and the temperature are state-dependent too. However, if $\xi$ depends \textit{linearly} on $r_h$, adopting tilde coordinates ensures that $T$ remains state-independent. On the other hand, if $\xi$ is state-independent, then we do not need to use tilde variables to maintain this property.

More generally, for any state-dependent $\xi$ there exists an appropriate rescaling such that $T$ becomes state-independent, for any function $f$ of the above type. This is the essential conclusion of this discussion.

We now introduce the most general 2d dilaton gravity action, in absence of additional matter or gauge fields, \cite{Grumiller:2021cwg}
\eq{
I[e_a,\omega,X,X^a] = \frac{k}{2\pi}\int\Big(X\,\extd\omega + X^a\,\big(\extd e_a + \epsilon_a{}^b\,\omega\wedge e_b\big) + \frac12\,\epsilon^{ab}\,e_a\wedge e_b\, {\cal V}(X,\,X^c X_c)\Big) \,.
}{eq:ana1}
Here, $k=1/(4G)$ is the gravitational coupling constant, $e^a$ is the zweibein 1-form, $\omega$ is the dualized Lorentz connection 1-form, $X$ is the dilaton field, and $X^a$ are auxiliary scalar fields. Latin indices are raised and lowered with the Minkowski metric, which we always put into lightcone gauge, $\eta_{+-}=1$, $\eta_{\pm\pm}=0$. We fix the sign in the antisymmetric tensor such that $\epsilon^\pm{}_\pm=\pm 1$.

The theory defined by the action \eqref{eq:ana1} is classically equivalent to a theory defined by the second-order action
\eq{
I[g_{\mu\nu},\,X] =- \frac{k}{4\pi}\,\int\extd^2x\sqrt{-g}\,\Big(XR-2{\cal V}\big(X,\,-(\partial X)^2\big)\Big) \,,
}{eq:ana2}
with the same potential function $\cal V$.

Since one of the most important points, to associate a dilaton model to an analogue system, is to establish whether a given quantity is state-dependent or state-independent, we now summarize various quantities in Table \ref{tab:lengthdim} , some of which will be introduced later, and indicate their state-dependence. As we adopt certain non-standard dimensional commonly used in analogue gravity literature, we also introduce in the Table the corresponding length dimensions for clarity.

\begin{table}[h]
\centering
\begin{tabular}{|c|l|}
  \hline
  dim & quantity \\ \hline
  - 2 & ${\color{blue}{\xi}}$, $T$, ${\color{red}{E}}$, $R$, $\cal V$;  \\
  - 1 & ${\color{blue}{\lambda}}$; \\
  0 & ${\color{blue}{k}}$, ${\color{blue}{r}}$, $r_h$, $f$, ${\color{red}{S}}$, ${X}$, $X_{0}$, $\tilde X$, $U$, $V$, $Z$, $Q$, $Q_0$;  \\
  2 &  ${\color{blue}{v}}$, $\extd s^2$, ${\color{blue}{\extd x^2}}\sqrt{-g}$. \\
  \hline
\end{tabular}
\caption{Dimensionalities of key quantities used throughout this paper, together with their dependence on the system’s state. Quantities in red are always \textit{state-dependent}, those in blue are always  \textit{state-independent}, and those in black may have either behavior.}
\label{tab:lengthdim}
\end{table}

This Table captures the state-dependence of quantities used in generic dilaton gravity theory. Our proposed procedure however, imposes certain constraints, which change the nature of some of the quantities.

We are now ready to start our journey.

\section{Solving the equations of motion and more assumptions}\label{sec:EOM}

The equations of motions descending from \eqref{eq:ana1} are
\begin{subequations}\label{EOM1storder}
\begin{align}
   & \extd X+ X^a \epsilon_a{}^be_b=0 \label{equation for Xa} \,, \\
   & \extd X^a- X_b\epsilon^{ba}\omega+\epsilon^{ab}e_b \,{\cal V}=0 \,, \\
   & \extd\omega+\frac{1}2\epsilon^{ab}e_a\wedge e_b \,\partial_X {\cal V}=0 \label{EOM 1} \,,\\
   & \extd e_a + \epsilon_a{}^b\,\omega\wedge e_b+\frac{1}2\epsilon^{cb}e_c\wedge e_b\, \partial_{X^a} {\cal V}=0 \label{equation torsion} \,,
\end{align}
\end{subequations}
for $\delta \omega, \delta e^a, \delta X$ and $\delta X^a$, respectively.  They are solved explicitly in \cite{Grumiller:2021cwg}. Here we report the formulas most relevant for us, especially focusing on minor differences with \cite{Grumiller:2021cwg}, due to our choice in Table \ref{tab:lengthdim}.

We define the dimensionless quantity
\eq{
\tilde X = \frac{X^+X^-}{\lambda^2X^2} \,,
}{eq:ana4}
and relabel the potential as
\eq{
{\cal V}(X,\,2X^+X^-) = \lambda^2\, V(X,\,\tilde X)\,.
}{eq:ana5}

The conservation equation
\eq{
\extd\tilde X + \bigg(2\tilde X - \frac{V(X,\,\tilde X)}{X}\bigg)\,\frac{\extd X}{X} = 0 \,,
}{eq:ana6}
in principle can be integrated to yield the dilaton $X$ as function of $\tilde X$, or vice versa. \textcolor{black}{The integration constant $X_{0}$, stemming from there, will play an important role later on.} The precise relation depends on the specific class of models and will be made clear later, once we know the relevant class.

Also the auxiliary function $Q$ defined by
\eq{
 \frac{\extd Q}{\extd\tilde X}=-\frac{1}{X^2}\,\frac{\partial V}{\partial X} \,\frac{\extd X}{\extd\tilde X} \,,
}{eq:ana7}
can be integrated to a function of $\tilde X$, once we express $X$ as a function of $\tilde X$.

The arbitrary integration constant $Q_0$ will be fixed suitably below. It can be either state-dependent or state-independent, depending on the model. Note that $Q_0$ is additive in $Q$ and therefore multiplicative in $e^Q$.

The general expression obtained for the metric, in the dilaton language, is then \cite{Grumiller:2021cwg}
\eq{
\extd s^2 = 2 e^Q \extd v\extd X - \lambda^2 \, 2 \tilde X X^2 e^{2Q(\tilde X)} \,\extd v^2 \,,
}{eq:dilaMetricGen}
which we require to be the metric \eqref{eq:ana3}
\[
\extd s^2 = 2 \extd v\extd r - \xi \, f(r - r_h)\,\extd v^2 \,.
\]

To achieve this, we first identify
\eq{
\xi \equiv\lambda^2 \,,
}
{eq:ana8bis}
and
\eq{
f(r-r_h) \equiv 2\tilde X X^2 e^{2Q(\tilde X)} \,,
}{eq:ana9}
where $X^2 (\tilde X)$, i.e. $X$ is expressed as function of $\tilde X$, using the results above (see later \eqref{eq:ana16} for an expression holding in the cases of interest here).

Notice that by virtue of $\xi$ being a tunable lab parameter, in \eqref{eq:ana8bis} we have related a certain phenomenological parameter to a dilatonic one. The same is true for the equation \eqref{eq:ana9}, where we rewrite the analog metric function in terms of dilaton coordinates.

Then we ask the radial coordinate to satisfy
\eq{
dr \equiv e^{Q(\tilde X)} dX = e^{Q(\tilde X)} \frac{dX}{d \tilde X} d \tilde X  \;,
}
{eq:anadrdX}
which also reads
\eq{
\frac{dr}{d \tilde X} = e^{Q(\tilde X)} \frac{dX}{d \tilde X} \;.
}
{eq:anadr/dtildeX}
Integrating, and requiring that the horizon, $r_h$, is reached when $\tilde X = 0$, we have
\eq{
r-r_h = \int\limits^{\tilde X}_0 e^{Q(\tilde X')}\,\frac{\extd X}{\extd\tilde X'}\,\extd\tilde X' \,.
}{eq:ana8}
The latter request is justified as follows. The location of the Killing horizon is determined by the zero of the function $f(r-r_h)$, hence, assuming $Q$ is finite and $X^2$ is positive, a Killing horizon must correspond to a locus where $\tilde X=0$. This is made explicit in \eqref{eq:ana8}, by adjusting the lower integration limit.

On the other hand, the assumption we made earlier, that $r_h$ is \textit{state-dependent}, goes hand in hand with the fact that $X_h$ (that corresponds to ${\tilde X} = 0$) is state-dependent.

\section{Dilaton scale invariant models}\label{sec:6}

For all dilaton models, the first law of BH thermodynamics holds
\eq{
\extd E = T\,\extd S \;,
}{eq:ana12}
and allows to determine the energy $E$ from temperature $T$ and entropy $S$. In general, one may view $T$ as a function of $S$, $T=T(S)$.

The typical BH analogue metrics available in laboratories have the form \eqref{eq:ana3}. In these setups, the temperature (essentially $\xi$) is operationally state-independent. As discussed at length earlier, in the models considered here, the state-dependance of $T$ can be turned on and off. Henceforth, we might request that the class of dilaton models that describe these analogue systems possesses this property.

The only way for this to hold, is that the models exhibit \textit{dilaton scale invariance}. Concretely, for any solution $X$ to the equations of motion, $\lambda X$ is also a solution of the same equations of motion, where $\lambda$ is an arbitrary positive number, which may be either state-independent or state-dependent.

In fact, one can always perform state-dependent diffeos that turn the state-dependence of $T$ on or off, independent of the model. However, in general the associated state-space has no obvious interpretation as some 2d dilaton gravity state-space. This has been shown explicitly for the Schwarzschild BH in
Section \ref{sec:preliminaries}. Dilaton scale invariant models are the only known exception \cite{Grumiller:2021cwg}, as there are always choices where the constant of motion disappears from the metric and only appears in the dilaton field (multiplicatively). Dilaton-scale invariant models thus generalize the discussion of JT gravity given in the Section \ref{sec:preliminaries}.

Working with such models gives us the freedom to scale the time coordinate and therefore to change the normalization of the timelike Killing vector. As discussed in Section \ref{sec:preliminaries} (see \eqref{tscale}), This allows one to change the operational meaning of temperature $T$. Therefore, we can always find a set of coordinates, where $T$ is state-independent. Similarly, one can always find a set of coordinates, where $T$ is state-dependent. The question of which of these coordinate choices is realized in the laboratory is a separate and non-trivial practical issue.

Together with the non-negotiable state-dependent nature of entropy, means that temperature, when this family of models is in the appropriate coordinates, does not depend on entropy, which in turn means that it does not depend on the horizon-value of the dilaton. In these special cases, then, the first law \eqref{eq:ana12} integrates trivially to
\eq{
E = T S \,.
}{eq:ana13}

We emphasize that even the same model, for example JT gravity, can lead to different integrated first laws. In particular, one may have $\partial_S T = 0$ (see the rescaled lapse function \eqref{FtildeJT}), leading to \eqref{eq:ana13}, or $\partial_S T \neq 0$, leading to a different result. Indeed, for the standard JT (see \eqref{FJT}), $T = a S$, with $a$ constant, that through \eqref{eq:ana12}, gives $dE = a S dS$, or $E = a \, S^2/2 =  T \, S/2$, that differs from \eqref{eq:ana13}. In Section \ref{sec:JT} we comment further on this point.

Dilaton scale invariance also implies a restriction \cite{Grumiller:2002md}
\eq{
V(X,\,\tilde X) = X\,U(\tilde X),
}{eq:ana14}
which considerably simplifies the potential, since it is only a function of one argument.

The decisive property of potentials \eqref{eq:ana14} is that they map the actions \eqref{eq:ana1} and \eqref{eq:ana2} to themselves, times an overall multiplicative factor under rescalings of the dilaton.

\section{Output from input}\label{sec:masterODE}
\subsection{The dilaton potential}%
\label{sub:The dilaton potential}

The remaining task is to relate the function $U(\tilde X)$, that due to the scale invariance depends only on a single variable, to the input function $f(r-r_h)$.

Defining the new function
\eq{
Z(\tilde X) \equiv \int_0^{\tilde X}\frac{\extd y}{U(y)-2y}
}{eq:ana15}
with $Z(0) = 0$, allows to solve the conservation equation \eqref{eq:ana6} for $X$
\eq{
X = X_0\,e^{Z(\tilde X)}.
}{eq:ana16}

The knowledge of $Z(\tilde X)$ allows to recover the desired potential $U(\tilde X)$ through
\eq{
U(\tilde X) = 2\tilde X + \frac{1}{Z'(\tilde X)}\,.
}{eq:ana17}
So our task can be rephrased as finding a solution for the function $Z(\tilde X)$, which which we proceed to do.

Plugging the potential \eqref{eq:ana14} and the dilaton solution \eqref{eq:ana16} into the differential equation \eqref{eq:ana7} allows to solve the latter in terms of the function $Z$
\eq{
Q(\tilde X)=Q_0 + \ln Z^\prime(\tilde X) - 2Z(\tilde X) \,,
}{eq:ana18}
hence
\eq{
e^{Q(\tilde X)} = e^{Q_0}\,Z'(\tilde X)\,e^{-2Z(\tilde X)} \,.
}{eq:ana18bis}

\subsection{Themodynamics, Ricci scalar and various behaviors for analogue models}%
\label{sub:Themodynamics, Ricci scalar and various behaviors for analogue models}

Let us now consider, the Hawking temperature, as given in \eqref{eq:Tana}. Substituting the identifications \eqref{eq:ana8bis} and \eqref{eq:ana9} into the geometric expression yields
\eq{
T=\frac{\xi}{4\pi}\,\partial_r f(r-r_h)\big|_{r=r_h} \equiv \frac{\lambda^2}{2\pi}\,X^2 e^{2Q(\tilde X)}\,\frac{\extd\tilde X}{\extd r}\bigg|_{\tilde X=0} \,,
}{eq:ana10}
which expresses $T$ in terms of dilaton quantities on the far right side.

Using \eqref{eq:ana16}, \eqref{eq:ana18bis} and \eqref{eq:anadr/dtildeX} in \eqref{eq:ana10}, we obtain
\eq{
T = \frac{\lambda^2}{2\pi} \, \left( e^{Q_0}X_0 \right)  \,,
}{eq:anaTeQ0X0}

As required, $T$ must be state-independent. However, with the exception of $\lambda$, the quantities in \eqref{eq:ana16} may exhibit state-dependence. Thus, to find additional constraints, we utilize the fact that entropy and energy are, by definition, state-dependent quantities.

We first consider the (Wald) entropy, which, for any dilaton model, is proportional to the dilaton evaluated at the horizon:
\eq{S = k\,X\big|_{\tilde X=0} \,,
}{eq:ana11}
and using \eqref{eq:ana16} yields
\eq{S = k \, X_0 \,.
}{eq:ana23}
Therefore, if $T$ is state-independent, we can employ the integrated first law \eqref{eq:ana13}:
\eq{TS = \frac{k\lambda^2}{2\pi}\,X_0^2e^{Q_0} \qquad\Rightarrow\qquad E=\frac{k\lambda^2}{2\pi}\,X_0^2 e^{Q_0} \,.
}{eq:ana24}
This firmly establishes that $X_{0}$ is a state-dependent quantity, given that neither $k$ nor $\lambda$ are state-dependent. To eliminate the state-dependence of $T$, we impose the condition
\eq{e^{Q_0} X_0 \equiv 1 \,,
}{eq:anaQ0X0}

which implies that $Q_{0}$ is state-dependent.

As discussed extensively, for the scale invariant dilaton models, that we identified as the most viable dilaton models, we have the freedom to turn on and off the state dependence of $T$. In cases where $T$ is state dependent, such as the standard JT solution (see \eqref{FJT}), the constraint \eqref{eq:anaQ0X0} is no longer needed, hence $Q_0$ can also be state-independent.

In conclusion, for a state independent $T$, we obtain
\begin{equation} \label{eq:label}
   T= \frac{\lambda^{2}}{2\pi}, \quad E = \frac{k \lambda^{2}}{2 \pi} X_{0},
\end{equation}
which demonstrate that energy is linear in $X_{0}$ and that any physical state can be labeled by its value.

\subsection{The Master equation}%
\label{sub:The Master equation}

Next, we consider the radial coordinate in \eqref{eq:ana8}. Inserting \eqref{eq:ana16}, \eqref{eq:ana18bis} and \eqref{eq:anaQ0X0} into \eqref{eq:ana8} yields
\eq{
r-r_h = \int\limits_0^{\tilde X}(Z'(y))^2e^{-Z(y)}\extd y\,.
}{eq:ana19}
On the other hand, applying the inverse function of $f$ to the crucial relation \eqref{eq:ana9} establishes a different equality for $r-r_h$, viz.
\eq{
r-r_h = f^{-1}\Big[2\tilde X \big(Z'(\tilde X)\big)^2\,e^{-2Z(\tilde X)}\Big]\,.
}{eq:ana20}
Equating the right hand sides of \eqref{eq:ana19} and \eqref{eq:ana20} yields an integro-differential equation that we convert into a second order ODE by differentiating once with respect to $\tilde X$
\eq{
\big(Z'(\tilde X)\big)^2\,e^{-Z(\tilde X)} = \frac{\extd}{\extd\tilde X}\, f^{-1}\Big[2\tilde X \big(Z'(\tilde X)\big)^2\,e^{-2Z(\tilde X)}\Big] \,.
}{eq:ana21}
For any input function $f$ the master ODE \eqref{eq:ana21} can be solved for $Z(\tilde X)$ (either numerically or, for judicious choices of $f$, in closed form), which by virtue of \eqref{eq:ana17} produces the desired output function $U(\tilde X)$.

A shorter version of the master ODE above is obtained by substituting
\eq{
y=e^{-Z} \,.
}{eq:ye-Z}
Note that this implies $y\in[0,1]$ with $y\to 1^-$ corresponding to approaching the horizon from the outside and $y\to 0^+$ corresponding to approaching the asymptotic region. With this the master ODE \eqref{eq:ana21} reduces to
\eq{
\frac{(y')^2}{y} = \frac{\extd}{\extd\tilde X} f^{-1}\big[2\tilde X (y')^2\big] \,.
}{eq:ana21too}


\section{Ricci scalar and various behaviors for analogue models}\label{sec:thermo}

%
%
%
Geometrically, the Ricci scalar associated with metrics \eqref{eq:ana3} is given by
\eq{
R = - \xi \,f''(r-r_h)\,.
}{eq:R1}
On the other hand, the Ricci scalar for a dilaton scale-invariant model with some given function $U(\tilde X)$ is given by \cite{Grumiller:2002md}
\eq{
R=2U-4UU'+4\tilde X (U')^2+8\tilde X^2U''-4\tilde X UU''\,.
}{eq:R2}

This yields the correct results for JT ($U= - \xi$), $R= - 2 \xi$ and Witten BH ($U=\tilde X + \lambda$), $R=2(\tilde X-\lambda)\propto M$.

Assuming that the relation between the radial coordinate $r$ and the dilaton $X$ is such that $r\to\infty$ implies $X\to\infty$ , table \ref{tab:1} shows the behavior of various functions at infinity, at the horizon, and in the centre.

The best-defined locus is the horizon, where the first five entries in table \ref{tab:1} have precisely defined numerical values, and the remaining three are still better defined than at any other locus. This suggests using the horizon as a starting point for an expansion.

\begin{table}[H]
\centering
    \begin{tabular}{l|l|l|l}
    Function & Asymptotic & Horizon & Centre \\\hline
    $X$ & $+\infty$ & $X_0>0$ & $\underline{0}$ \\
    $\tilde X$ & $<0$ & $0$ & $>0$ \\
    $Z$ & $+\infty$ & $0$ & $\underline{-\infty}$ \\
    $y$ & $0$ & $1$ & $\underline{+\infty}$ \\
    $\epsilon$ & $1$ & $0$ & $\underline{-\infty}$ \\
    $U$ & \underline{finite} & finite and nonzero & unknown\\
    $Q$ & unknown & finite & unknown \\
    $Z^\prime$ & unknown & finite and nonzero & unknown
    \end{tabular}
    \caption{Behavior at infinity, horizon, and centre of various key quantities. The underlined values are typical values, i.e., in exceptional cases they can be different from what reported here.}
    \label{tab:1}
\end{table}

\section{Theory next to experiments}\label{sec:results}

In this Section we eventually use the machinery we have put together step by step, bearing in mind that, in particular, one of the conclusions of the previous Sections is that the dilaton models suited for analogue gravity systems are the scale invariant models.

With this framework established, on the one hand, we aim to explore whether known and important gravity models, useful for the experimental exploration of important theoretical problems, are within the reach of laboratory realizations (top-down approach). On the other hand, we seek to identify what dilaton gravity models are realized by the typical analogue BHs \cite{BHchipNature2023,PachosTanhPRL.130.016701,Linear12018,Linear2Pachos2023} (bottom-up approach). Let us start with the latter.

\subsection{Bottom-up approach}

The phenomenological set-up we focus on first is the linear approximation of the typical $f = \tanh(r -r_h)$ BH profiles \cite{BHchipNature2023,PachosTanhPRL.130.016701}, that are also discussed separately in \cite{Linear12018,Linear2Pachos2023}, where $f=\gamma \;(r-r_{h})$. For a linear function the inverse function is also linear and \eqref{eq:ana21too} simplifies to
\eq{
\frac{(y')^2}{y} = \frac{1}{\gamma} \frac{\extd}{\extd\tilde X} \big[2\tilde X (y')^2\big],
}{eq:ana21togamma}
or equivalently to
\eq{
y'' = \frac{y'}{2 \tilde X} \left(\frac{\gamma}{2 y}-1\right).
}{eq:ana21togamma2}
Even in this simplified form, we can not find a solution in closed form, thus we shall solve the equation perturbatively, using a near-horizon expansion
\eq{
y=1-\epsilon \,.
}{eq:y1eps}
With this, the equation takes the form
\eq{
\epsilon''= \frac{\epsilon'}{4\tilde X} \frac{\gamma +2 \epsilon-2}{1-\epsilon} \,.
}{eq:epsilongamma}

We present the first four terms, as this will be sufficient for certain key considerations. The solution, at fourth order,  can be expressed as
\begin{align}
    \epsilon(\tilde{X}) &= -\alpha^2 \tilde{X}-\frac{\alpha^4  \tilde{X}^2}{2 (\gamma -6)}-\frac{\gamma  \alpha^6 \tilde{X}^3 \left(\gamma+12 \right)}{6 (\gamma -10) (\gamma -6)} \nonumber\\
    &-\frac{ \alpha^8  \tilde{X}^4 \gamma \left(\gamma^3+54  \gamma ^2-72  \gamma -2160  \right)}{24 (\gamma -14) (\gamma -10) (\gamma -6)^2}+{\cal O}(\tilde X^{5}) \,,
\end{align}
where  $\alpha^2$ is a constant fixed by the boundary condition on $y^{'}(0)$. Using $Z(\tilde X)=-\ln(1-\epsilon(\tilde X))$ and \eqref{eq:ana17}, $U(\tilde X)=2\tilde X + 1/Z'(\tilde X)$, the near-horizon expansion of the potential $U(\tilde X)$ is
\begin{gather}
    U(\tilde{X})=\frac{-1}{\alpha^2}+\frac{2 \gamma-6}{\gamma -6}\tilde{X}+\frac{\alpha^2 \left(5 \gamma ^2-6 \gamma \right) \tilde{X}^2}{(\gamma -10) (\gamma -6)^2}-\frac{12 \alpha^4 \left(5 \gamma ^2+12 \gamma -180\right) \tilde{X}^3}{(\gamma -10) (\gamma -6)^3}+{\cal O}(\tilde X^{4}) \,.
\end{gather}
As can be seen from the geometric expression for $R$ in \eqref{eq:R1}, the linearity of $f$ implies a Ricci flat metric. On the dilaton side, then, the potential should lead to the vanishing of\footnote{Numerical tests up to the 42th order show that this condition holds till higher orders.} \eqref{eq:R2}
\begin{gather}
   R=\frac{\gamma -2}{\gamma -6} \Big{(} \frac{6 }{\alpha^2}+\frac{4 \left(\gamma ^2+3 \gamma+90\right) \tilde X}{(\gamma -10) (\gamma -6)}\Big{)}+{\cal O}(\tilde X^{2}) =0 \,.
\end{gather}
This condition constraints $\gamma$ to be 2, therefore \eqref{eq:epsilongamma} reduces to
\eq{
\epsilon'' + \frac{\epsilon'}{2\tilde X} = \frac{\epsilon'}{2\tilde X(1-\epsilon)} \,.
}{eq:ana42}
The fact that $T = \xi \gamma /(4 \pi)$ is not in contradiction with this arbitrariness as we do not have any unit for the temperature, so even fixing $\gamma$ to the specific value 2, does not fix the actual, phenomenological value of the temperature.
Another consideration is that $U(\tilde{X})$ is related to $Z'(\tilde{X})$ (see above), thus a shift in $Z$ as $Z \to Z+\delta$ would not affect the discussions above. However, one should note that we have assumed the horizon to be located at $Z=0$. Shifting in $Z$ will alter the location of the horizon and our assumption. Thus, when this location is fixed, we can not change it anymore.

\begin{figure}
\begin{center}
\includegraphics[width=0.5\linewidth]{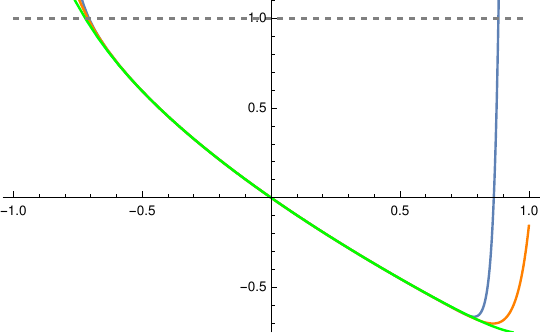}
\caption{Perturbative solution for $\epsilon(\tilde X)$ with $\alpha=1$. Blue: 42 Taylor coefficients. Orange: 20 Taylor coefficients. Green: 10 Taylor coefficients.}
\label{fig:1}
\end{center}
\end{figure}

We now move further in the Taylor expansion to address other issues. In Fig.\ref{fig:1} we plot the behavior of $\epsilon (\tilde X)$ for 10, 20 and 42 terms of the expansion, and the corresponding expressions for $U(\tilde{X})$ follow as before.

In the asymptotic region it is a bit more convenient to use $y$
\eq{
y'' = \frac{y'}{2\tilde X}\,\Big(\frac1y-1\Big)\,.
}{eq:ana21three}
Since asymptotically we get $y\to0^+$ and $\tilde X\to\tilde X_{\textrm{\tiny asy}}$ we expand \eqref{eq:ana21three}
\eq{
y'' = \frac{y'}{2\tilde X_{\textrm{\tiny asy}}\,y}\,\big(1+{\cal O}(y)+{\cal O}(\tilde X-\tilde X_{\textrm{\tiny asy}})\big)
}{eq:ana44}
and, neglecting the terms that vanish asymptotically, we find an \textit{exact solution} for $y$ in terms of two integration constants. One of them is fixed by consistency, i.e., $y\to 0^+$ as $\tilde X\to \tilde X_{\textrm{\tiny asy}}^+$. The other one, denoted below by $c_1$, is free but can be fixed by demanding consistency with the $y\to 1$ at $\tilde X\to 0$, i.e., $y$ takes the correct value at the horizon. Of course, this is not necessarily a good requirement since the exact solution of \eqref{eq:ana44} is not supposed to be accurate near the horizon. Nevertheless, it turns out that the difference between the asymptotic solution and the near horizon expansion remains relatively small everywhere. In Mathematica notation, the function is given by $y = (H^{-1}) (\tilde{X}- \tilde{X}_{asy})$ where
\eq{
H(x) = 2 e^{-c_1} \tilde{X}_{asy} \, {\rm Ei}(c_1 + \ln x).
}{eq:ana45}
%

Close to $\tilde X_{\textrm{\tiny asy}}$ the qualitative difference between the $\epsilon(\tilde X)$ near horizon and its asymptotic behavior is significant: while the Taylor expansion has finite values for all its derivatives, the asymptotic solution has a pole in $\epsilon''$ and in $Z'$ at $\tilde X_{\textrm{\tiny asy}}$.

\begin{figure}
\begin{center}
\includegraphics[width=0.5\linewidth]{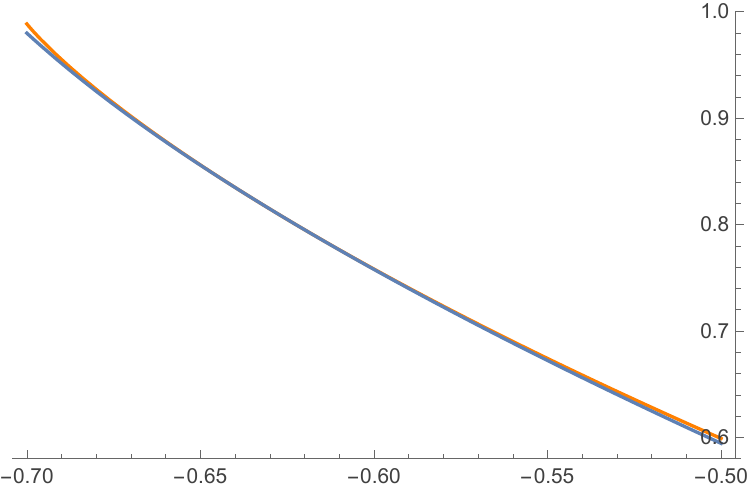}
\caption{Plots of $\epsilon$ between asymptotic region and horizon. Orange: asymptotic solution, blue: near horizon Taylor expansion with 42 coefficients.}
\label{fig:2}
\end{center}
\end{figure}

Numerical analysis using 42 Taylor coefficients show that the match of the solutions such that $\epsilon$ and $\epsilon'$ are continuous there, is possible. In Fig. \ref{fig:2} the comparison between the curves for a match at $\tilde X = 1/2$ is shown.  This fixes the two integration constants in the asymptotic solution and allows to determine more precisely the value of $\tilde X_{\textrm{\tiny asy}}$.

Moreover, the asymptotic solution has the following properties:
\begin{itemize}
    \item By construction, the asymptotic value is $\epsilon(\tilde X_{\textrm{\tiny asy}})=1$.
    \item The first derivative has a log-branch cut, $\epsilon^\prime \propto \ln(\tilde X-\tilde X_{\textrm{\tiny asy}}) + \dots$ with finite prefactor.
    \item The second derivative has a first-order pole, $\epsilon''\propto 1/(\tilde X-\tilde X_{\textrm{\tiny asy}})+ \dots$ with finite residue.
\end{itemize}
This behavior suggests a multi-logs function as the solution. At next to leading order (NLO), we use
\eq{
    \epsilon = 1-(\tilde X-\tilde X_{\textrm{\tiny asy}}) \,
    \frac{\ln(\tilde X-\tilde X_{\textrm{\tiny asy}}) + \ln(-\ln(\tilde X-\tilde X_{\textrm{\tiny asy}}) +\dots)}{2\tilde X_{\textrm{\tiny asy}}}
    + {\cal O}(\tilde X -\tilde X_{\textrm{\tiny asy}})^2 \,,
}{eq:epsNLO}
which, qualitatively and also quantitatively, captures all the key features of $\epsilon$ mentioned above. Moreover, the result shows that the residue of the pole in $\epsilon''$ is numerically the same as the prefactor of the log-term in $\epsilon'$ and both are given by $-1/(2\tilde X_{\textrm{\tiny asy}})$, which is close to the numerical results.

Based on the NLO solution for $\epsilon$ we find the asymptotic result
\eq{
    e^{-Z} = \frac{X_0}{X} = \frac{(\tilde X-\tilde X_{\textrm{\tiny asy}})\ln(\tilde X-\tilde X_{\textrm{\tiny asy}})}{2\tilde X_{\textrm{\tiny asy}}}\,\big(1+\dots\big) \,,
}{eq:eZNLO}
    where the ellipsis denotes terms that vanish at least like $\ln(-\ln(\tilde X-\tilde X_{\textrm{\tiny asy}}))/\ln(\tilde X-\tilde X_{\textrm{\tiny asy}})$. The solution for $\tilde X$ as a function of the dilaton $X$ is
\eq{
    \tilde X = \tilde X_{\textrm{\tiny asy}} + \exp\bigg[\texttt{ProductLog}\Big[-1, \frac{2\tilde X_{\textrm{\tiny asy}} X_0}{X}\Big]\bigg] \,,
}{eq:tildeXNLO}
which asymptotically expands as
\eq{
    \tilde X = \tilde X_{\textrm{\tiny asy}} +\frac{2\tilde X_{\textrm{\tiny asy}} X_0}{X\,\ln\frac{-2\tilde X_{\textrm{\tiny asy}} X_0}{X}}\,\big(1+{\cal O}(\ln\ln X/\ln X)\big) \,.
}{eq:tileXNLO2}
This yields the asymptotic form of the potential function $U(\tilde X)$:
\eq{
    U = 2\tilde X + \frac{1}{Z'} = \tilde X + \tilde X_{\textrm{\tiny asy}} + \frac{\tilde X -\tilde X_{\textrm{\tiny asy}}}{\ln(\tilde X-\tilde X_{\textrm{\tiny asy}})} + \dots \,.
}{eq:UNLO}
Thus, the first two terms are precisely as for the Witten BH and the third term gives the leading contribution to $U''$ (which vanishes for the Witten BH). Nonetheless, there is no known dilaton model these analogue systems, with linear $f \propto r - r_h $, realize.

We can now turn our attention to the general typical case, see, e.g., \cite{BHchipNature2023,PachosTanhPRL.130.016701}, that is $f = \tanh(r-r_h)$. That means that the inverse function is the arctanh, thus its derivative is $\textrm{arctanh}'(x)=1/(1-x^2)$. The master ODE \eqref{eq:ana21too} is then
\eq{
y'' + \frac{y'}{2\tilde X}\,\Big[1 - \frac{1}{2y}\Big] + \frac{{\tilde X} {y'}^5 }{y} = 0 \,.
}{eq:ana21tanh}
The main difference with the linear case is the last term, which becomes dominant asymptotically ($y \to 0$). The near-horizon behaviour ($y \to 1, \tilde X \to 0$), essentially, coincides with the linear function result in \eqref{eq:ana21togamma2} (here, though, $\gamma = 1$) and thus captures the near-horizon behavior of the function $y$. Indeed, the corresponding  near horizon Taylor expansion, up to second order, matches the linear case. Indeed, there is something generic about all BH states, in the sense that all of them need to have a single non-degenerate zero, so there is always a regime where the function $f$ is well-approximated by a linear function. On the other hand, the asymptotic expansion can be rather different, as we now show.

In fact, the equation in the asymptotic regime contains an extra term
\eq{
y'' - \frac{y'}{4 {\tilde X}_{asy} \, y}\, + \frac{{\tilde X}_{asy} {y'}^5 }{y} = 0 \,,
}{eq:ana21tanhASY}
that makes the analytic solution (in this approximate regime) even more cumbersome than the one we found for the linear case.

The numerical analysis to match the two regimes, then, gives similarly complicated results and the final expression for the dilaton potential, $U$, results, at least analytically, in a unclassified dilaton BH. Once more, we could not identify a useful dilaton model corresponding to this analogue system.

In conclusion, the two important cases just analyzed, corresponding to the typical 2d analogue BH and its linear approximation, show that if we want to simulate dilaton gravity BHs and their thermodynamical properties, we need to step-up the laboratory techniques to realize different devices and architectures.

\subsection{Top-down approach}

In this approach we start with known forms of $Z(\tilde X)$ and then reverse-engineer the associated input function $f(r-r_h)$ (that actually becomes the \textit{output}). Let us begin with
\eq{
Z (\tilde X) = \tilde X \,,
}{eq:zx}
corresponding to the potential $U=2\tilde X+1$ or, equivalently, ${\cal V} = X + 2X\tilde X$.
Within the so-called $ab$-family of models \cite{grumi3}, this model corresponds to the choices $a=2$ and $b=-1$.

Inserting $Z=\tilde X$ into \eqref{eq:ana20} yields
\eq{
f(r-r_h) = 2\tilde X e^{-2\tilde X}\,.
}{eq:ana36}
Using \eqref{eq:ana19} yields $\tilde X=-\ln[1 - (r-r_h)]$ and thus
\eq{
f(r-r_h) = - 2 \ln[1 - (r-r_h)][1 - (r-r_h)]^2\,.
}{eq:ana37}
The function \eqref{eq:ana37} is a function of $(r - r_h)$ and it has a single zero at $r=r_h$, as desired for our analogue models. Furthermore, the temperature is state-independent
\[
T = \frac{\xi}{4\pi}\,\partial_r f(r - r_h)\big|_{r=r_h} = \frac{\xi}{2\pi} \,,
\]
as wanted.

This example immediately demonstrates that reverse-engineering the function $f(r-r_h)$, from a given potential function $Z$, is the appropriate approach, if our goal is to realize tractable and interesting theoretical models. Let us now present one more important example of this kind.

If we choose
\eq{
Z (\tilde X) = c \ln(\tilde X+1) \,,
}{eq:zx}
that also belongs to the $ab$-family of models, with $a=2+\frac1c$ and $b=-1-\frac1c$ \cite{grumi3}.

Two such models are particularly important: $c = - 1$ and $c = - 1/2$. Let us look at the first.

When $c=-1$, from \eqref{eq:ana20} one has $f(r - r_h) = 2 \tilde X$, while from \eqref{eq:ana19} we obtain
\eq{
f(r-r_h) = 2 \left( e^{r-r_h} - 1\right) \,.
}{eq:anafwitten}
This is the $f$ for the Witten BH metric in EF gauge \cite{grumi3}. As before, the function \eqref{eq:anafwitten} is a function of $(r - r_h)$, it has a single zero at $r=r_h$ and the temperature is
\[
T = \frac{\xi}{4\pi}\,\partial_r f(r - r_h)\big|_{r=r_h} = \frac{\xi}{2\pi} \,.
\]

The two cases just illustrated are promising theoretical avenues to pursue. However, the experimental challenges posed by the concrete realization of these $f(r - r_h)$s, lie beyond the scope of this paper.

When $c=-1/2$, we have $1/Z' = - 2 (\tilde X + 1)$, that in \eqref{eq:ana17} gives $U = - 2$. This is JT gravity. Both the Witten BH and the JT BH are of standard form and dilaton scale invariant, with $U(\tilde X)=\tilde X - 1$, the former, and $U(\tilde X)= - 2$, the latter. Nonetheless, if we proceed for the JT potential as we did for the previous two cases, we run into a contradiction, as we shall explain in the dedicated next Section.

\section{Lessons for JT gravity in the lab}\label{sec:JT}

If we consider JT BH, identified here by $U = - 2$, and proceed similarly to the Witten BH, we should utilize \eqref{eq:zx} with $c = - 1/2$ in \eqref{eq:ana20}. This leads to
\eq{
  f(r - r_h) = \frac{1}{2} \, \frac{\tilde X}{\tilde X + 1} \,,
}{eq:anafJTwrong}
Correspondingly, from \eqref{eq:ana19}, we obtain
\begin{eqnarray} \label{eq:anafwrongJT}
    f_{+}(r-r_h) & = & 2 \left( (r - r_h) - (r - r_h)^2 \right) \,,   \\
    f_{-}(r-r_h) & = & 1-2r+2\left( r+r^2 - 2rr_h+r_h^2 \right) \,.
\end{eqnarray}

The function $f_{+}$ has two real roots, while $f_{-}$ has two complex roots; therefore, only the roots of $f_{+}$ are physically relevant.

However, this does not reproduce the known form of the expression of $f$ for the JT BH which is given by
\eq{
f_{JT}(r, r_h) = 2 \left( r^2 - r^2_h \right) \,.
}{eq:anafJT}
In fact, the functions have different asymptotics, as $f_{JT}\propto r^{2}$ and $f_{+}\propto -r^{2}$. As a consequence, their associated scalar curvatures have opposite signs.

This discrepancy indicates that the formulae \eqref{eq:ana19} and \eqref{eq:ana20} cannot be directly applied. These relations rely on the assumption that $f(r, r_h)$ can be expressed as $f (r - r_h)$, a form justified by the properties of the analogue metrics considered in \cite{BHchipNature2023,PachosTanhPRL.130.016701,Linear12018,Linear2Pachos2023}. Since $f_{JT}(r,r_h)$ does not satisfy this form, these equations are not valid here. Additionally note that the temperature is state-dependent
\eq{
T_{JT} = \frac{\xi}{4\pi}\,\partial_r f(r - r_h)\big|_{r=r_h} = \frac{\xi}{2\pi} r_h \,,
}{eq:anaTJT}
henceforth $e^{Q_0} X_0 = 1$ does not hold and \eqref{eq:ana19} and \eqref{eq:ana20} strongly rely on the constraint \eqref{eq:anaQ0X0}.

This naturally brings us back to the discussion initiated in Section \ref{sec:preliminaries}, where we analyzed the state-dependence of $T_{JT}$. As explained there, the
state-dependence of $T_{JT}$, from the mathematical point of view, can be turned on and off, by selecting the appropriate choice of the coordinates, namely $(v,r)$ and $(\tilde v, \tilde r)$. Henceforth, in principle, we could remove the state-dependence from $T_{JT}$,  thus enabling the application of \eqref{eq:ana19} and \eqref{eq:ana20}.

Nonetheless, as discussed in Sections \ref{sec:preliminaries} and \ref{sec:setup}, the operational meaning of the temperature $T$ ultimately depends on the physical measurement setup. If the relevant physical quantity has the same nature of the $\xi$ of the qubit/couplers transmons system described in \cite{BHchipNature2023}, then the laboratory coordinates are of the $(\tilde v, \tilde r)$ type. This means that we have a state-independent $T_{JT}$, and one part of the problem is solved. Nonetheless, the actual realization of a quantum circuit capable of reproducing the function $f$ remains a highly non-trivial task. Another key unresolved issue concerns the operational interpretation of the time variable, which represents a rescaled lab time, $\tilde v = r_h \, v$.

It is important to emphasize once more that the thermodynamics of the JT BH depends crucially on this choice of coordinate. In the frame $(v,r)$, $T_{JT} = a S$, while in the frame $(\tilde v, \tilde r)$, $T_{JT}$ is just a constant number, hence independent from $S$.

In summary, the questions of how to realize JT gravity experimentally, remains open. The standard approaches require a deep rethinking that accounts for the considerations presented in this paper. We have outlined the key theoretical issues involved in designing physical systems that could emulate this gravity model.

\section{Conclusions and outlook}

Given the growing need for the experimental settling of key theoretical disputes of quantum and classical gravity, we deem important that the theoretical community resorts more and more to analogue experiments and to the use of gravity/field theory correspondences. In other to do that, we need to know the advantages and limitations of such an approach, to pave the way to a correct and mathematically sound use of those experimental set-ups.

In particular, for the 2d systems realized in experiments, and for those important for the theory (JT/SYK), the subtleties and peculiarities of the 2d need to be fully exploited. This is what we do here, focusing on the typical 2d analogue BHs, $f(r,r_h) = \tanh(r-r_h)$, as those recently realized in the lab with quantum superconductive circuits \cite{BHchipNature2023} or studied in theoretical investigations of phase transitions of spin-chains \cite{PachosTanhPRL.130.016701}. Our goal has been to first set-up a list of assumptions and then to identify the dilaton model corresponding to those analogue BHs. We could not find any analytic solution to the problem, hence performed extensive numerical calculations, which reveals that the analogue BHs analyzed here do not correspond to interesting and important dilaton gravity models. This severely limits the use of such systems to solve open problems on the theory side.

We have followed a step by step procedure to carefully identify the most reasonable physical and mathematical assumptions on the dilaton model that should correspond to the given analogue BH. The most important assumption, stemming from the physical systems under scrutiny, is that the temperature, $T$, of those analogue BHs is state-independent. This means two things: the value of $T$ does not change when the location of the horizon, $r_h$, changes, that is an instance that can be verified in the lab; $T$ and $S$ can be made independent quantities, something that leads to scale-invariant dilaton models. Other assumptions are obtained through physical requests, like, e.g., $r_h$ being state-dependent, or mathematical ones, like the nature of the zeroes of $f(r,r_h)$ and that $f(r,r_h) = f(r - r_h)$.

This latter assumption that $f(r,r_h) = f(r - r_h)$ is motivated entirely by the analogue systems realized in the labs, and it appears a rather restrictive requirement from an experimental perspective (see, e.g., \cite{BHchipNature2023}). On the other hand, if analogue experimental setups were able instead to implement functions of the type
\begin{equation}\label{last}
        f(r,r_h) = f_1(r) (f_2(r) - f_2(r_h)) \,,
\end{equation}
then the location of the Killing horizons would still be fixed at $r = r_h$ but such functions would accommodate the ``standard'' 2d dilaton gravity models, with the potential of the form ${\cal V} = U(X) X^+ X^- + V(X)$, while simultaneously allowing for a description with state-independent temperature. Throughout this work we have taken care to respect what current analogue experiments are capable of realizing, but it is also part of our task to lay out the most natural theoretical directions and leave the corresponding challenges to the experimental community. Thus, the last part of the paper is devoted to reverse the logic to what we called ``top-down'' approach, focusing instead on the ``standard'' models of \eqref{last}, such as JT, Witten BH or CGHS. In these cases it is immediate to obtain the $f(r,r_h)$s the analogue systems must realize, whereas the difficulty shifts to designing experimental set-ups, such as quantum circuits, that can implement these specific profiles.

One may go even further and consider the most general case of a Killing norm of the form
\begin{equation}\label{lastlast}
        f(r,r_h) \,,
\end{equation}
with \textit{two independent arguments}. This would probably map to the most general 2d dilaton gravity action \eqref{eq:ana2} and the two arguments of $f$ translate into corresponding two arguments of $\cal V$. However, this mapping, in general, is not bijective in our analysis we have already encountered distinct models that lead to the same form $f = (r - r_h)$.

A central challenge for experiments, as becomes clear here, is the full handling of state-dependence/state-independence of $T$. For instance, although JT gravity can be formulated with a state-independent $T$, this requires working in the operational frame where such a description is valid. Conversely, it remains an intriguing issue whether a state-dependent $T$, corresponding to the usual JT gravity, can be faithfully realized in the laboratory.

These challenges may prove demanding, but the potential payoff is substantial. Successful implementations of the corresponding analogue systems would address important theoretical disputes and long‑standing theoretical questions could finally be examined under controlled experimental conditions.

\section*{Acknowledgement}
We are all much indebted to Daniel Grumiller, who  was the driving force of this research in the early days of the paper.

PC, AI and JK, gladly acknowledge the financial support of Charles University Grant UNCE 24/SCI/016. MSN's work is based upon research funded by Iran National Science Foundation (INSF) under project No.4037031. MSN also acknowledges the financial support received from the Federal Ministry of Education, Science, and Research, OeAD-GmbH Agency for International Mobility and Cooperation in Education, Science, and Research.

\end{document}